\documentstyle[psfig,ijmpc]{article}

\begin{document}
\normalsize\textlineskip
{\thispagestyle{empty}
\setcounter{page}{1}
 
\renewcommand{\thefootnote}{\fnsymbol{footnote}} 
\def\bsc{{\sc a\kern-6.4pt\sc a\kern-6.4pt\sc a}}
\def\bflatex{\bf L\kern-.30em\raise.3ex\hbox{\bsc}\kern-.14em
T\kern-.1667em\lower.7ex\hbox{E}\kern-.125em X}
 
\copyrightheading{Vol. 0, No. 0 (1997) 000--000}
 
\vspace*{0.88truein}

\def\OR{\hbox{\vrule height 6pt depth -3pt \vrule height 1pt depth 2pt}
\hskip 2pt}

\def \mtrx{{\mathbf{\widetilde{M}}} }
\def \graph{{\mathcal{G}}}
\def \crit{{\mathcal{C}}}
\def \back{{\mathcal{B}}}
\def \dir{\mathcal{D}}
\def \dirofg{{\dir}(\graph)}
\def \graphab{{\graph}_{ab}}
\def \dirofgab{{\dir}(\graphab)}
\def \critab{{\crit}(a,b)}
\def \backab{{\back}(a,b)}
\def \visab{{\mathcal{V}}(a,b)}
\def \queue{{\mathcal{Q}} }

\fpage{1}
\centerline{\bf \large A fast algorithm for backbones}
\vspace{0.37truein}
\centerline{\footnotesize C.~Moukarzel}
\vspace*{0.015truein}
\centerline{\footnotesize\it  Instituto de F\'{\i}sica, Universidade
Federal Fluminense}
\baselineskip=10pt
\centerline{\footnotesize\it 24210-340 Niter\'oi, Rio de Janeiro , Brazil }
\vglue 10pt
\centerline{\footnotesize\it  e-mail: cristian@if.uff.br}
\vglue 10pt
\publisher{(received date)}{(revised date)}
 
\vspace*{0.21truein}

\abstracts{\noindent A matching algorithm for the identification of
backbones in percolation problems is introduced. Using this procedure,
percolation backbones are studied in two- to five-dimensional systems
containing $1.7\times10^7$ sites, two orders of magnitude larger than was
previously possible using burning algorithms. }{}{}

\vspace*{1pt}\textlineskip
\section{Introduction}
\vspace*{-0.5pt}
\noindent
\textheight=7.8truein
\setcounter{footnote}{0}
\renewcommand{\thefootnote}{\alph{footnote}}

In percolation models~\cite{percolation}, the sites or bonds of a lattice
are independently occupied with probability $p$, and the properties of the
resulting clusters are studied. Above a critical density $p_c$, a cluster
that spans the whole system exists, and one says that connectivity
percolates. The backbone is defined as the subset of a connected cluster
that carries current when a potential difference is applied between two
points. The backbone of the spanning cluster determines the macroscopic
transport properties of the system. At the critical concentration $p_c$,
connected clusters and backbones are known to be fractals, although with
different fractal dimensions. While geometrical properties of connected
clusters have been numerically studied~\cite{HK} on very large systems
containing as much as $10^{11}$ sites, backbone studies~\cite{burning} have
been reduced to relatively small lattices due to the lack of efficient
integer algorithms for backbone identification. The burning algorithm has
been used by several authors, but this procedure is strongly CPU-limited and
therefore it has not been possible to study systems of more than $2.5 \times
10^5$ sites.

Matching algorithms have been recently introduced~\cite{BHalg,CFMalg,JHalg}
for the related problem of rigidity percolation~\cite{letters,JT}, and I
show here that an extension of these methods can be applied to the study of
connectivity percolation, allowing very efficient backbone identification
for systems much larger than was previously possible using the burning
algorithm.

\section{The matching algorithm}

Consider a system of $n$ sites $i=1,\ldots,n$ connected by an arbitrary set
$E$ of $b$ bonds $ij$ connecting sites $i$ and $j$. Bonds are initially
labelled from 1 to $b$, and sites are labelled from $b+1$ to $b+n$. The
matching algorithm can be thought of as a clever way to identify and remove
loops, and is implemented using a directed graph $\dir$ as an auxiliary
representation of the system. In this directed graph, lattice sites are
represented by nodes $i$ and bonds by directed edges $ij$. We may think of
each edge as an arrow. These can be pointing in either direction, subject to
the constraint that \emph{no node has more than one arrow pointing to it}. A
node pointed to by an arrow will be said to be \emph{covered} by the
corresponding edge. A node with no incoming arrows is on the other hand 
\emph{uncovered}.
\begin{figure}[t]
\centerline{\psfig{figure=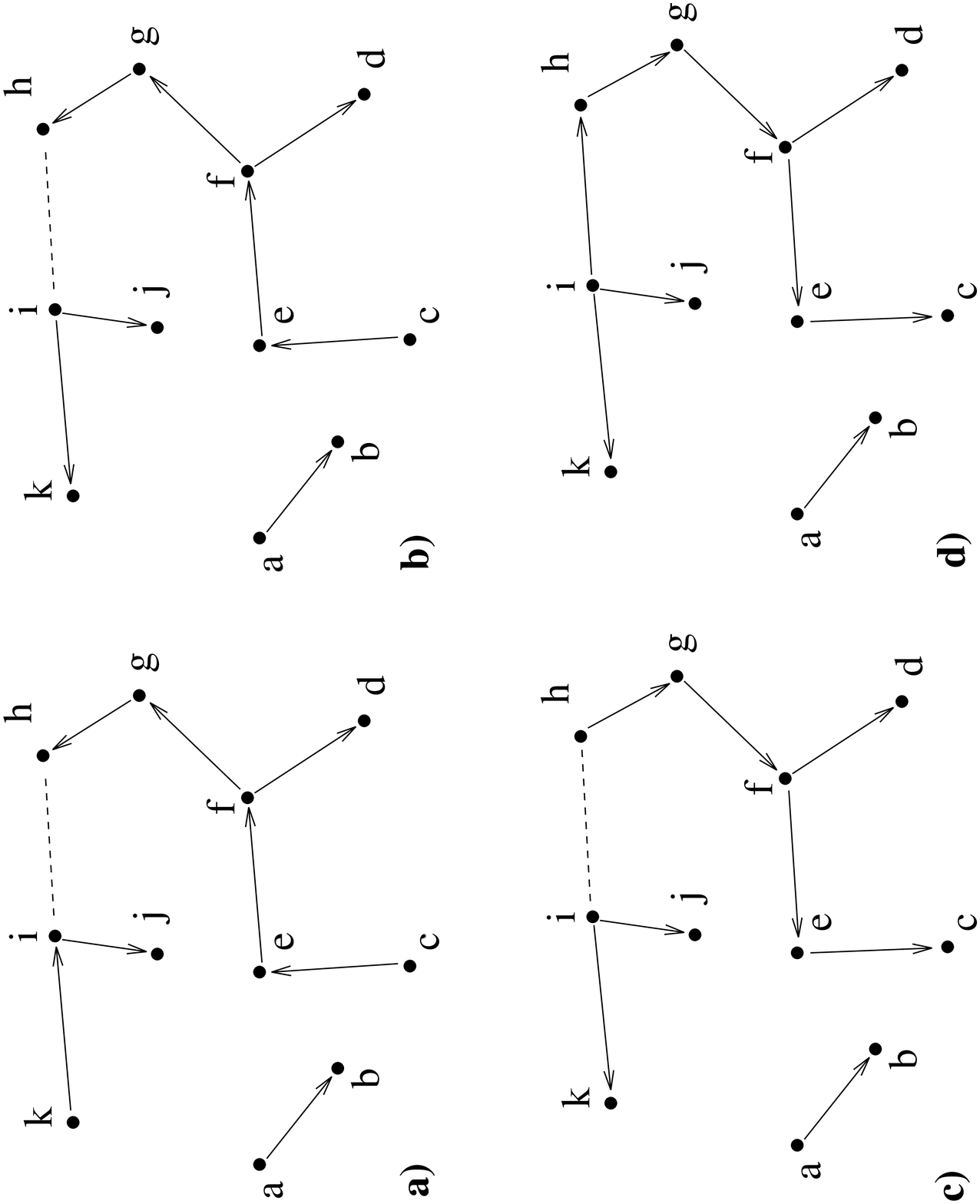,width=8cm,angle=270}} 

\caption{ {\bf a)} Before adding a new edge (ih) (dashed line) the following
test is done: {\bf b)} arrow (ki) is inverted, uncovering i. {\bf c)} we
uncover h by inverting the entire path (hgfe). Since both ends could be
simultaneously uncovered, the new edge does not close a loop and therefore
{\bf d)} it is definitively added to \protect $\dir$.  }

\label{fig:matching-succeeds} 
\end{figure} 
The directed graph $\dir$ will be kept loopless. In order to do this, each
time a closed loop (a cycle, or circuit) is identified (see later) it will
be removed from $\dir$, and replaced by a loop-node. A loop-node is a node
in $\dir$ that represents a deleted loop. Therefore, although initially
there are as many nodes in $\dir$ as there are sites in our system, as the
algorithm proceeds we will delete some nodes and create loop-nodes.
These loop-nodes will be given a loop-label, which is the minimum of all
edge- and node-labels in the loop.

We start from a graph $\dir$ initially containing $n$ nodes and no edges,
and add edges one at a time. Before adding an edge $ab \in E$ to $\dir$, the
following test is done in order to know if a loop is closed by $ab$: We
attempt to reorganise the existing arrows (if any) on $\dir$, in order to
uncover both $a$ and $b$ simultaneously. Since by hypothesis $\dir$ without
edge $ab$ has no loops, it must always be possible to uncover one of them.
Let us then assume that $a$ is uncovered first. If after doing this $b$ can
also be uncovered (while keeping $a$ uncovered), then the new edge $ab$ does
not close a loop, and therefore it is definitively added to $\dir$. This
means: we add an arrow between $a$ and $b$, pointing to either of them
(Fig.~\ref{fig:matching-succeeds}).

If on the other hand it is not possible to uncover $b$ while keeping $a$
uncovered (Fig.~\ref{fig:matching-fails}), this necessarily means that a
loop would be formed on $\dir$ by the addition of $ab$. In this case, edge
$ab$ is \emph{not} added to $\dir$ but the following is done instead:
Starting from $b$ (covered), we follow the arrows \emph{backwards}. This
will necessarily lead to $a$, thus identifying the new loop. All edges in
this loop are given a common label $l_{min}$, which is the minimum amongst
all node- and edge-labels in the loop (including nodes $a$ and $b$ and edge
$ab$). Next all nodes and edges in the loop are removed from $\dir$ and
replaced by a node with label $l_{min}$.

\begin{figure}[t]
\centerline{\psfig{figure=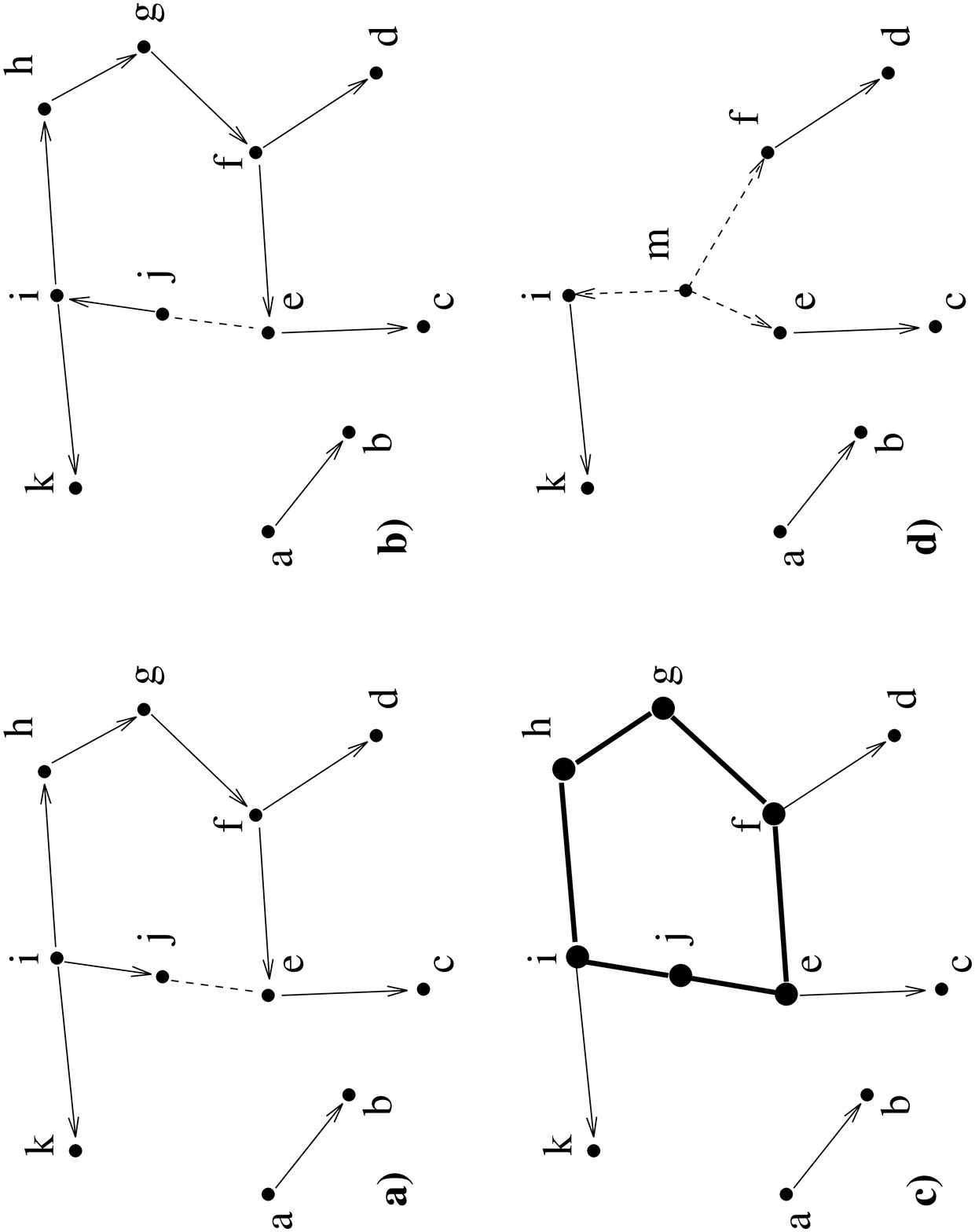,width=8cm,angle=270}} 

\caption{ {\bf a)} We attempt to add a new edge (je) (dashed) to \protect
$\dir$. {\bf b)} First arrow (ij) is inverted, uncovering j. But now there
is no way to uncover e while keeping j uncovered. This means that (ej)
closes a loop. So now: {\bf c)} start from e and follow arrows backwards, in
this way identifying all edges and nodes in the loop (bold lines).  {\bf d)}
The loop is removed from \protect $\dir$ and replaced by a loop-node m. }

\label{fig:matching-fails} 
\end{figure} 

There is an exception though, as eventually some of the nodes in this loop
are connected to other nodes outside the loop (for example nodes i, f and g
in Fig.~\ref{fig:matching-fails}). These are not removed from $\dir$ but
connected instead to the newly created node (node m in
Fig.~\ref{fig:matching-fails}) by auxiliary arrows that initially point
outwards. These auxiliary arrows have the same label as the loop-node. Also,
some of the nodes that we do remove (nodes j, g and h in
Fig.~\ref{fig:matching-fails}) may be eventually needed at later stages, for
example if we have to add a new edge that connects to one of them. In such a
case we simply recreate the corresponding node and connect it by means of an
auxiliary arrow to the loop node.

The process of replacing identified loops by nodes is called
\emph{condensation}~\cite{CFMalg}. It is possible to implement the matching
algorithm without condensation. In this case, one simply does not add edges
that close a loop but simply keeps track of the fact that a new loop has
been formed by giving all loop-edges a common label. The algorithm is
perhaps simpler in this way. Condensation has on the other hand the effect
of enormously improving speed and reducing memory requirements, since the
graph size is decreased each time a loop is found. This is extremely
important for large systems~\cite{CFMalg}.

\section{Backbone Identification}

Imagine we want to know if two far apart points $a$ and $b$ on the system
are connected, and in such case we want to identify the backbone $\backab$
between these two points. This can be done by noticing that, if $a$ and $b$
are connected, then a long-range bond $ab$ would close a loop. Thus our
method above serves to identify the parts of the system that belong to this
loop, i.e. the backbone $\backab$.

\begin{figure}[t] 
\centerline{\psfig{figure=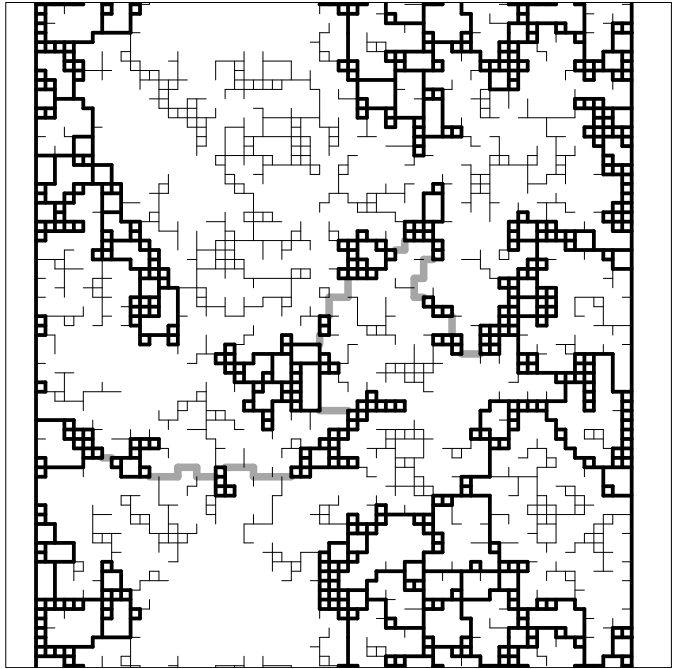,width=9cm}}
\caption{A spanning cluster on a site-diluted square lattice of linear
dimension $L=64$ at the critical point. Bus bars are located on the left and
right ends of the sample. Boundary conditions are periodic in the vertical
direction. The removal of any critical bond (thick gray lines) would produce
disconnection. In addition to those, `blobs' of multiply connected bonds
(thick black lines) also belong to the backbone. Dangling ends (thin lines)
are connected to the backbone at one point only and are not relevant for
macroscopic conductivity.}
\label{fig:a-backbone}
\end{figure}

Therefore at any point we may simply do as if we were to connect a
fictitious edge $ab$ between two arbitrary points $a$ and $b$ in the form
described above, that is: we attempt to uncover both $a$ and $b$ on $\dir$.
If this is possible, then $a$ and $b$ are not connected. End. If only $a$
but not $b$ can be uncovered, starting from $b$ follow the arrows backwards
and in this way the backbone between $a$ and $b$ is identified. The visited
edges are all critical, i.e. cutting one of the would produce disconnection
of $(a,b)$. These are called \emph{red bonds} in the percolation literature.
Some of the visited nodes will be loop-nodes. These are also included in the
backbone, i.e. all edges that form the loops are, as well as all loop-nodes
in them, in a recursive manner. Loop nodes are what is called \emph{blobs}
in the percolation literature. They are multiply connected so they contain
no red bonds. If after identifying the backbone one is interested in also
identifying the dangling ends, this can be done by `testing' a fictitious
bond between the backbone and all other nodes in the system. If a node is
connected to the backbone then one cannot uncover both ends simultaneously.
Thus the whole spanning cluster can be identified in this way
(Fig.~\ref{fig:a-backbone}). But let us stress that dangling-end
identification is not one of the strengths of the matching algorithm. This
would be more efficiently done using for example a Hoshen-Kopelmann type
algorithm~\cite{HK}.

This is  just a brief sketch of the main ideas about the matching algorithm.
Several important issues have not been discussed in this paper, as for
example how exactly nodes are uncovered in an efficient manner, data
structures needed for computational speed, as well as a precise discussion
of the relevant properties of the algorithm. A more detailed description
will be published elsewhere~\cite{CFMtbp}.

\section{Results}

We present here partial results for  scalar percolation on site-diluted
hypercubic lattices in 2 to 5 dimensions. We start with a system containing
no sites and add them one at a time at random locations. Each time a site is
occupied, all new induced edges are tested and added to $\dir$ one at a
time. Induced edges are those connecting the new site to already occupied
nearest-neighbours. Loops are identified and removed as described in the
previous sections. We proceed in this way until the percolation point is
reached, which is detected because of the existence of a fictitious
long-range bond connecting opposite sides of the sample. At this point, the
density of occupied sites gives $p_c$ for this sample. We identify and
measure the whole backbone (red bonds plus blobs) as well as the dangling
ends at $p_c$ for each sample in this manner.

\subsection{CPU time}

Figure~\ref{fig:cpu} shows CPU times per sample in seconds, needed to
identify the whole backbone as well as the spanning cluster at $p_c$ versus
linear sample size. Approximately half of that time is needed to identify
the backbone alone, and the rest is used to obtain the dangling ends.
Largest sizes simulated were $L=4096$ in 2D, $L=256$ in 3D, $L=60$ in 4D and
$L=26$ in 5D. The number of samples varies between some thousands for the
largest sizes to some millions for the smaller ones in each dimension. All
runs were done on Alpha-500 workstations. Fits of the data in
Fig.~\ref{fig:cpu} show that CPU-time scales with system size $n$ as $t \sim
n^\theta$ with $\theta=$ 1.07 (2D), 1.06 (3D), 1.05 (4D) and 1.05 (5D).

\begin{figure}[t]
\centerline{\psfig{figure=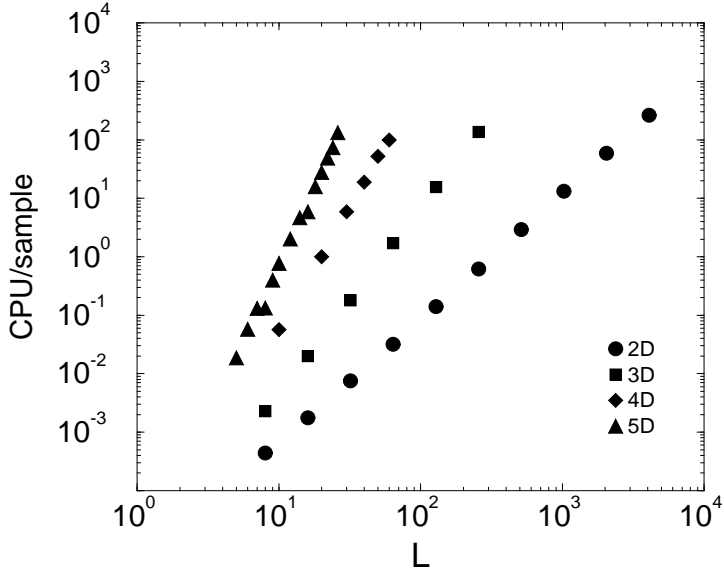,width=10cm}}
\caption{ CPU time $t$, in seconds per sample versus linear size $L$, needed
to identify the whole spanning cluster of scalar connectivity at $p_c$.
Identifying the backbone alone would take half of that approximately. The
time-complexity exponent $\theta$ in $t \sim n^\theta = L^{d \theta}$ is
approximately one in all cases.}
\label{fig:cpu}
\end{figure}

\subsection{Correlation length exponent \protect $\nu$}

Since we add sites one at a time until the percolation point is reached, we
are able to measure, for each sample, the critical density $p_c$ of occupied
sites. The fluctuation $\sigma_L = <p_c^2>_L-<p_c>_L^2$ of this quantity,
were $<>_L$ indicates averages overs samples of size $L$, is expected to
scale as $\sigma_L \sim L^{-1/\nu}$ with system size~\cite{percolation}. Here
$\nu$ is the correlation length exponent. On the other hand, rigorous
arguments due to Coniglio~\cite{Coniglio} show that the number $R_L$ of red
bonds grows at $p_c$ as $L^{1/\nu}$.

\begin{figure}[t]
\centerline{\psfig{figure=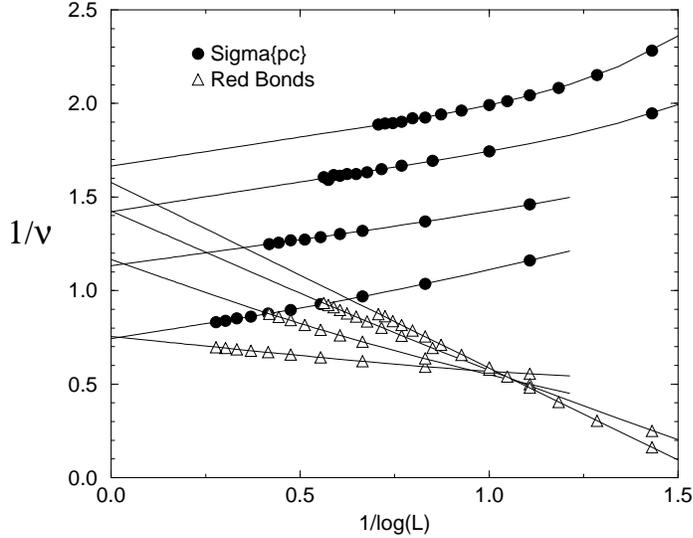,width=10cm}}
\caption{ Fits of \protect $-\log(\sigma_L)/\log(L)$ and \protect
$\log(R_L)/\log(L)$ versus \protect $1/\log(L)$ for scalar percolation in 2
to 5 dimensions. The intercept at \protect $1/\log(L)=0$ is the estimated
value of \protect $1/\nu$. }
\label{fig:nu-scalar-all}
\end{figure}

Thus measuring $\sigma_L$ and $R_L$, two independent estimates for the
thermal exponent $1/\nu$ are obtained. Figure~\ref{fig:nu-scalar-all} shows
plots of $-\log(\sigma_L)/\log(L)$ and $\log(R_L)/\log(L)$ versus
$1/\log(L)$. We fit these data using

\begin{equation}
\frac{1}{\sigma_L} , R_l \sim L^{1/\nu} \left( 1 + a L^{-\omega} \right)
\label{eq:nu-fit}
\end{equation}

\noindent where $\omega$ is a  corrections-to-scaling exponent. This allows
the following estimates: $1/\nu= 0.75 \pm 0.01$ (2D), $1.13 \pm 0.02$ (3D),
$1.44 \pm 0.05$ (4D) and $1.63 \pm 0.05$ (5D).

Correction to scaling exponents were found in most cases to be between 0.5
and 0.8, but precise values cannot be given for these lattice sizes. \\

\subsection{Backbone density at \protect $p_c$}

The fraction $B(L)$ of bonds on the backbone at $p_c$ is measured for
several system sizes $L$ in each dimension. Our results are shown in
Fig.~\ref{fig:bk-dens-scalar-all}.

\begin{figure}[t]
\centerline{\psfig{figure=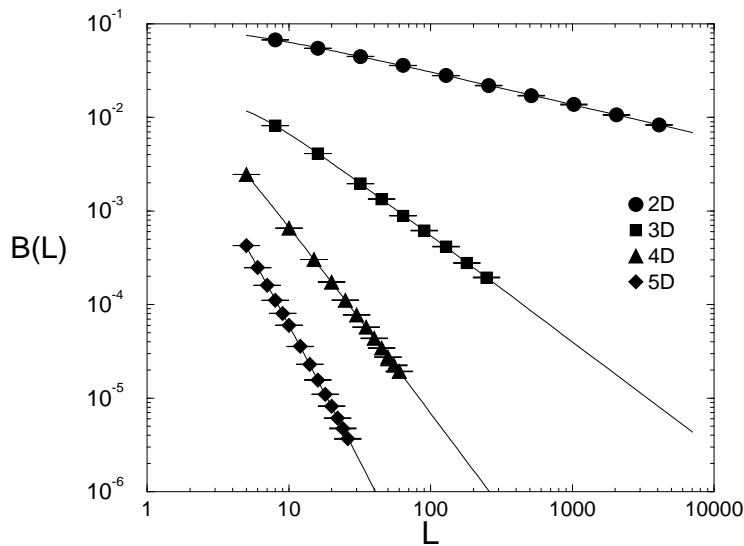,width=10cm}}
\caption{Backbone densities at $p_c$ for scalar percolation in 2 to 5
dimensions. Lines correspond to best fits using (\protect
\ref{eq:backbone-fit}).}
\label{fig:bk-dens-scalar-all}
\end{figure}

Now assume $B(L)$ to behave as

\begin{equation}
B(L) = \lambda L^{- \beta'/\nu} ( 1 + a L^{-\omega}  )
\label{eq:backbone-fit}
\end{equation}

\noindent where $d_b = d - \beta'/\nu$ is the backbone fractal dimension,
and $\omega$ is an exponent of corrections to scaling. Fitting this
expression with 4 free parameters to our data, we obtain the following
estimates for the backbone fractal dimension: $d_b = 1.650 \pm 0.005$ (2D),
$1.86 \pm 0.01$ (3D), $1.95 \pm 0.05$ (4D) and $2.00 \pm 0.05$ (5D).

\vskip 0.5cm
\leftline{\bf Acknowledgements} 

\noindent I whish to thank the kind hospitality of the Phys/Ast. Dept. of
Michigan State University, where this work was initiated, and Phillip
Duxbury for many motivating discussions. Support from the Conselho Nacional
de Desenvolvimento Cient\'\i fico e Tecnol\'ogico, CNPq, Brazil is also
acknowledged.

%
%
\eject

\end{document}